\documentclass[11pt]{article}
\usepackage{spconf,amsmath,epsfig}
\usepackage{times}
\usepackage{times}
\usepackage{color}
\usepackage{amsfonts}
\usepackage{amssymb}
\usepackage{amsbsy} % Per a \boldsymbol
\usepackage[ansinew]{inputenc}
\usepackage{pstricks}
\usepackage{lscape}
\usepackage{cite}
\usepackage{setspace}
\usepackage{rotating}
\usepackage{multirow}
\usepackage[hidelinks]{hyperref}
\usepackage{multicol}
\usepackage{float}
\usepackage{booktabs,caption,fixltx2e}
\usepackage{epstopdf}
\usepackage{upgreek}
\newcommand{\x}{{\mathbf x}}
\newcommand{\f}{{\mathbf f}}
\newcommand{\IG}{\includegraphics}

\title{Statistical Learning for End-to-End Simulations}
\name{J. Vicent$^*$, J. Verrelst$^*$, J.P. Rivera-Caicedo$^\dagger$, N. Sabater$^*$, J. Mu\~noz-Mar\'i$^*$, G. Camps-Valls$^*$ and J. Moreno$^*$.
\thanks{\bf Preprint, paper published in IGARSS 2018 - 2018 IEEE International Geoscience and Remote Sensing Symposium, Valencia, 2018, pp. 1699-1702, doi: 10.1109/IGARSS.2018.8518366.}
\thanks{This project was carried out in the frame of ESA's project {\it FLEX L2 End-to-End Simulator Development and Mission Performance Assessment} ESA Contract No. 4000119707/17/NL/MP. Jochem Verrelst was supported by the European Research Council (ERC) under the ERC-2017-STG SENTIFLEX project (grant agreement 755617).  Jordi Mu\~noz-Mar\'i was support by MINECO/ERDF under Grant CICYT TIN2015-64210-R. Gustau Camps-Valls was supported by the ERC under the ERC-CoG-2014 SEDAL project (grant agreement 647423). Contact: jorge.vicent@uv.es}}
\address{$^*$Image Processing Laboratory, University of Valencia, 46980 Paterna (Valencia), Spain. Web: http://ipl.uv.es.\\
 $^\dagger$CONACYT-UAN, Departamento: Secretaria de investigaci\'on y posgrado, Tepic, Nayarit, Mexico.}

\begin{document}

\maketitle

\begin{abstract}
End-to-end mission performance simulators (E2ES) are suitable tools to accelerate satellite mission development from concet to deployment. One core element of these E2ES is the generation of synthetic scenes that are observed by the various instruments of an Earth Observation mission. The generation of these scenes rely on Radiative Transfer Models (RTM) for the simulation of light interaction with the Earth surface and atmosphere. However, the execution of advanced RTMs is impractical due to their large computation burden. Classical interpolation and statistical emulation methods of pre-computed Look-Up Tables (LUT) are therefore common practice to generate synthetic scenes in a reasonable time. This work evaluates the accuracy and computation cost of interpolation and emulation methods to sample the input LUT variable space. The results on MONDTRAN-based top-of-atmosphere radiance data show that Gaussian Process emulators produced more accurate output spectra than linear interpolation at a fraction of its time. It is concluded that emulation can function as a fast and more accurate alternative to interpolation for LUT parameter space sampling.
\end{abstract}

\begin{keywords}
FLEX, sun-induced chlorophyll fluorescence, mission simulator, hyperspectral, Sentinel-3
\end{keywords} 

\section{Introduction}
\label{sec:intro}
The preparation of new satellite missions rely on simulated data to translate mission goals into technical requirements. In this respect end-to-end mission performance simulators (E2ES) become suitable tools to accelerate mission development from concept to deployment \cite{Verstraete2015419}. E2ES models both mission aspects (i.e., biogeophysical processes) and technical characteristics of platform and instruments (e.g., platform orbit and attitude) \cite{kerekes1}. On of the critical aspects of an E2ES is the generation of synthetic scenes that are to be onserved by the mission instruments. These scenes consist of spatially distributed hyperspectral Top-of-Atmosphere (TOA) radiance data, and should be described according to surface and atmospheric properties that are objective of a satellite mission \cite{Tenjo2017}. Radiative Transfer Models (RTM) emerge as a suitable option for scene generation given their representative simulation of the light interaction with the Earth surface and atmosphere. Since the computation cost of RTMs make them impractical for the simulation of large scenes, interpolation and emulation of pre-computed Look-Up Tables (LUT) are therefore common practice \cite{guanter2009,OHagan2006}.

The main objective of this work is to analyze the performance of emulators as an alternative of classical interpolation methods for a LUT TOA radiance data of based on MODTRAN \cite{Berk2006}. Our results show that emulation methods outperforms interpolation in both computational cost and accuracy, which suggest they might be better suited for RTM-based scene generation in satellite E2ES. The remainder of this paper is as follows: Section \ref{sec:theory} gives a theoretical overview of the analyzed interpolation and emulation methods. Section \ref{sec:methods} presents the materials and methods to study the performance of interpolation and emulation methods in terms of accuracy and computation time. This is followed by presenting the results in Section \ref{sec:results} which are discussed

\section{Interpolation and emulation theory}
\label{sec:theory}

Let us consider a $D$-dimensional input space $\mathcal{X}$ from where we sample $\x \in\mathcal{X}\subset \mathbb{R}^D$ in which a $K$-dimensional object function $\f(\x;\lambda) = [f(\x;\lambda_1),\ldots,f(\x;\lambda_K)]: \mathbb{R}\mapsto \mathbb{R}^K$ is evaluated. In the context of this paper, $\mathcal{X}$ comprises the $D$ input variables (e.g., Leaf Area Index (LAI), Aerosol Optical Thickness (AOT), Visual Zenith Angle (VZA)) that control the behavior of the function $\f(\x;\lambda)$, i.e., a water, canopy or atmospheric RTM. Here, $\lambda$ represents the wavelengths in the $K$-dimensional output space\footnote{For sake of simplicity, the wavelength dependency is omitted in the formulation in this paper, i.e., $\f(\x;\lambda)\equiv\f(\x)$.}.
An interpolation or an emulation, $\widehat\f(\x)$, is therefore a technique used to approximate model simulations, $\f(\x)=\widehat\f(\x)+\varepsilon$, based on the numerical or statistical analysis of an existing set of {\it nodes}, $\f_i=\f(\x_i)$, conforming a pre-computed LUT. Among all interpolation methods, {\bf piece-wise linear} interpolation is commonly used in remote sensing applications due to its balance between computation time and interpolation error ~\cite{guanter2009}. Its implementation is based on the Quickhull algorithm~\cite{Barber1996469} for triangulations in multi-dimensional input spaces. For scattered LUT input data, piece-wise linear interpolation is reduced to find the corresponding Delaunay simplex (e.g., a triangle when $D=2$) that encloses a query $D$-dimensional point $\x_q$:

\begin{equation} \label{eqn:lininterp}
\widehat{\f}_i(\x_q) = \sum_{j=1}^{D+1}\omega_j\f(\x_j) ,
\end{equation}

where $\omega_j$ are barycentric coordinates of $\x_q$ with respect to the $D$-dimensional simplex (with $D+1$ vertices).

An  emulator  essentially  functions  as  an  interpolation method,  but based  on  statistical  learning  principles. The  basic idea is that an emulator uses a limited number of simulator runs, i.e., input-output pairs (corresponding to training samples),  to  infer  the  values  of  the  complex  simulator  output given a yet-unseen input configuration. As with interpolation, the built emulator computes the output that is otherwise generated by the simulator \cite{OHagan2006}. Note that building an emulator is essentially nothing more than building a statistical learning regression model as often done for biophysical parameter retrieval applications, but in reversed order: whereas a retrieval model converts input spectral data (e.g., reflectance) into one or more output biophysical variables, an emulator converts input biophysical variables into output spectral data.

When  it  comes  to  emulating  spectral  outputs,  however, the challenge lies in delivering a full spectrum, i.e., predicting multiple spectral bands. It bears the consequence that the learning methods should be able to generate multiple outputs to be able reconstructing a full spectral profile.  This is not a trivial task.  Only few regression models can deal with multiple outputs.  However, training a complex multi-output statistical model with the capability to generate so many output bands would take considerable computational time and would probably incur in a certain risk of overfitting because of model over-representation.  A  workaround  solution  had  to  be  developed that enables the regression algorithms to cope with large, spectroscopy datasets.  An efficient solution is to take
advantage of the so-called curse of spectral redundancy , i.e., the Hughes phenomenon.  Since spectroscopy data typically shows a great deal of collinearity, it implies that such data can be  converted  to  a  lower-dimensional  space  through  dimensionality  reduction  techniques.   Accordingly,  by  converting
the spectral data into a limited set of components preserve the spectral information content it implies that the multi-output problem is greatly reduced \cite{Verrelst2017}.  Afterwards the components can then again be reconstructed to spectral data.

The first step thus involves building a statistically-based representation (i.e., an emulator) of the field data using statistical learning from a set of training data points derived from runs of the actual model under study (nodes in the context of interpolation).  These training data pairs should ideally cover the multidimensional input space using a space-filling algorithm.  The second step uses the emulator previously built in the first step to compute spectral output.  Based on the literature review above and earlier emulation evaluation studies \cite{Verrelst2017}, {\bf Gaussian processes regression} (GPR) is a powerful method to function as accurate emulators.

\section{Materials and Methods}
\label{sec:methods}

The intercomparison of interpolation and emulation methods is applied here on MODTRAN5 simulated TOA radiance spectra constructed according to (\ref{eqn:TOA}) under the Lambertian assumption:

\begin{equation}
    L_{TOA} = L_0 + \frac{(E_{dir}\mu_s+E_{dif})(T_{dif}+T_{dir}) \rho }{\pi(1-S \rho)} ,
\label{eqn:TOA}
\end{equation}

where $L_0$ is the path radiance, $E_{dir/dif}$ are the direct/diffuse at-surface solar irradiance, $T_{dir/dif}$ are the surface-to-sensor direct/diffuse atmospheric transmittance, $S$ is the spherical albedo, $\mu_s$ is the cosine of \emph{SZA}, and $\rho$ is the Lambertian surface reflectance (in our case we used a vegetation surface reflectance). 
A set of LUTs were generated by means of Latin hypercupe sampling (LHS) within the RTM variable space with minimum and maximum boundaries as given in Table \ref{table_inputs2}.  A LHS covers the full parameter space, ensuring that the interpolation/emulation reconstructs any possible combination of input variables. Three sizes of LUTs were created given the same boundaries: 500, 2000 and 5000. While the most dense LUT (5000) was used as a reference LUT to evaluate the performances of the emulation and interpolation algorithms, the first two LUTS (500 and 2000) where simulated to actually run the emulation and interpolation techniques. Additionally, the 64 vertex of the input variable space (i.e., where the input variables get the minimum/maximum values) were added to these two LUTs. The addition of these vertex enables consistent functioning of all tested interpolation techniques, i.e., that the input variable space is bounded and no extrapolation is performed.

\begin{table}[ht!]
\begin{center}
    \caption{Range of atmospheric LUT input variables distributed according to Latin Hypercube sampling. Fixed geometric variables: solar zenith angle: 55$^{\circ}$; observer zenith angle: 0$^{\circ}$; azimuth angle: 0$^{\circ}$.}
    \resizebox{0.50\textwidth}{!}{ 
     \begin{tabular}{llccc@{}}
\hline
       \multicolumn{2}{c}{\bf Model variables} & {\bf Units} & {\bf Minimum} & {\bf Maximum}\\
\hline			
O3C &  O$_3$ column concentration & [amt-cm] & 0.2 & 0.45 \\
CWV &  Columnar Water Vapour & scale-factor & 1 & 4 \\
AOT &  Aerosol Optical Thickness & unitless & 0.05 & 0.4 \\
G & Asymmetry parameter & unitless  & 0.65 & 0.99 \\
$\alpha$ &  \r{A}ngstr\"om exponent & unitless & 1 & 2 \\
SSA & Single Scattering Albedo & unitless & 0.75 & 1 \\
\hline
	   \end{tabular}}
	   \label{table_inputs2}
\end{center}
\end{table}

For the emulation approach, each LUT was used to develop and evaluate the GPR models.  Although the first 5 PCA components explained already 99.96\% of the variance, developing the regression model with less than 10 components led to inaccurate reconstructions. More precise profiles were achieved with having more components added in the regression algorithm. However, looping over more components slows down processing time. Therefore, 10 and 20 PCA components (i.e.,$\sim$100\% explained variance) were evaluated as a trade-off between accuracy and processing time. Further, since GPR emulators only produce an approximation of the original model, it is important to realize that such an approximation introduces a source of uncertainty referred to as ``code uncertainty'' associated to the emulator. Therefore, validation of the generated model is an important step in order to quantify the emulator's degree of accuracy. To test the accuracy of the 500- and 2000-LUT emulators, part of the original data is kept aside as validation dataset. In an attempt to keep processing fast, only a single split was applied using 70\% samples for training and the remaining 30\% for validation.

In order to inspect the emulation and interpolation accuracy some goodness-of-fit statistics are calculated against the $n$=5000 references LUTs as generated by the RTMs. The relative residuals (in absolute terms, expressed in \%) are calculated for each wavelength. Specifically, the average relative error and the percentiles 2.5\%, 16\%, 84\% and 95.5\% will be plotted. In addition, the root-mean-square-error (RMSE) and the normalized RMSE (NRMSE) [\%] (see (\ref{eqn:nrmse})) are calculated and averaged over all wavelengths ($\lambda$):

\begin{equation}
    NRMSE = 100\frac{\sqrt{n^{-1}\sum_{i=1}^n[\f(\x_i)-\hat\f(\x_i)]^2}}{\f_{max}-\f_{min}} , 
\label{eqn:nrmse}
\end{equation}

where $\f_{max}$ and $\f_{min}$ are respectively the maximum and minimum values of the $n$ spectra in the reference dataset. 

The processing time of executing the emulator/interpolation method on the reference dataset has also been tracked. These calculations were performed in a i7-4710MP CPU at 2.5GHz with 16 GB of RAM and 64-bits operating system.

\section{Results}
\label{sec:results}

In this section we will show the results of applying the emulator and interpolation methods on the described MODTRAN5 TOA radiance datasets. The results are given for the two training/interpolating LUTs (500 and 2000 samples). The emulation approach is additionally tested with entering 10 or 20 components in the regression algorithm. All approaches are validated against the reference 5000 samples' LUTs.

The averaged error statistics are shown in Table \ref{MODTRAN}. GPR is clearly overperforming the piece-wise linear interpolation method. The results indicate that GPR yielded RMSE$\boldsymbol{_\lambda}$ results more than 10 times lower than linear interpolation. Moving from 500 to 2000 training samples did not lead to significant improvements. The same holds for adding more components into the emulators, with a modest improvement in accuracy at expenses of an increase of computation time.
Regarding processing time, GPR emulation method overperforms agaist linear interpolation, delivering the 5000 spectra in less than 5 s in case of 2000LUT and trained with 20 components. Instead, linear interpolation method needs 1-3 minutes to reconstruct the 5000 reference spectra.

\begin{table}[!h] 
\centering 
 \caption{Interpolation and emulators validation results against 5000 LUT reference dataset.}
     \resizebox{0.48\textwidth}{!}{ 
 \begin{tabular}{lcccccc} \toprule %\toprule  
 {\bf Method} & \multicolumn{2}{c}{\bf RMSE$\boldsymbol{_\lambda}$} & \multicolumn{2}{c}{\bf NRMSE$\boldsymbol{_\lambda}$ (\%)} & \multicolumn{2}{c}{\bf CPU (s) } \\ 
 {\bf LUT training size:} &  {\bf 500} &  {\bf 2000} &  {\bf 500} &  {\bf 2000} &  {\bf 500} &  {\bf 2000} \\

\midrule%\midrule			
\multicolumn{2}{l}{\bf Linear interpolation:} &&&&& \\
  & 0.386 & 0.265 & 2.84 & 1.98  & 68 & 183 \\
{\bf GPR Emulation}: &&& &&&\\
 - 10PCA & 0.037 & 0.031 & 0.65 & 0.59 & 0.5 & 2.0 \\
 - 20PCA & 0.029 & 0.022 & 0.43 & 0.37 & 1.2 & 4.7 \\
\bottomrule
  \end{tabular}}
  \label{MODTRAN}
%\end{center}
\end{table}

\begin{figure}[h!]
	\centering
	\IG[width=\linewidth]{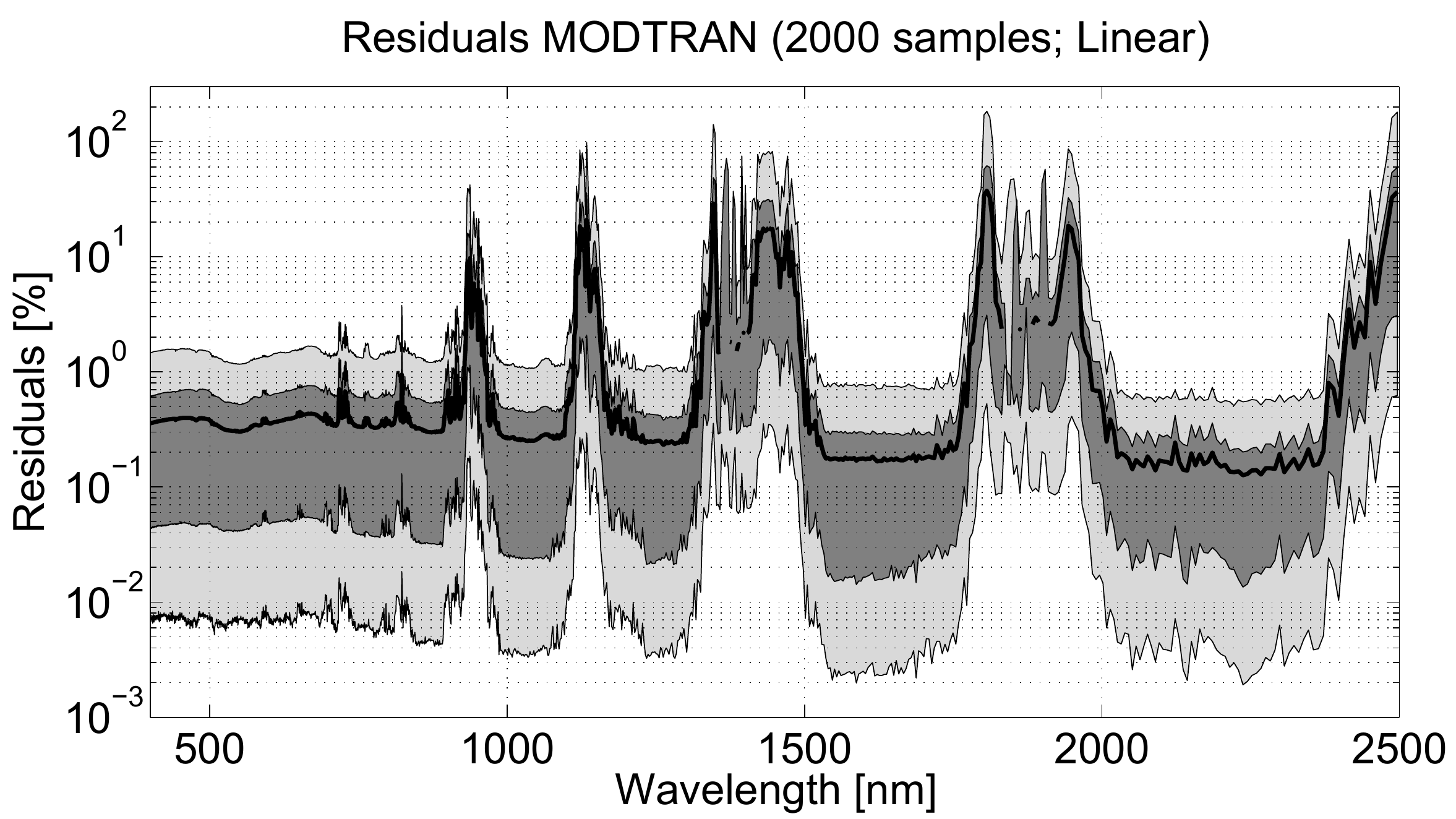}
	\IG[width=\linewidth]{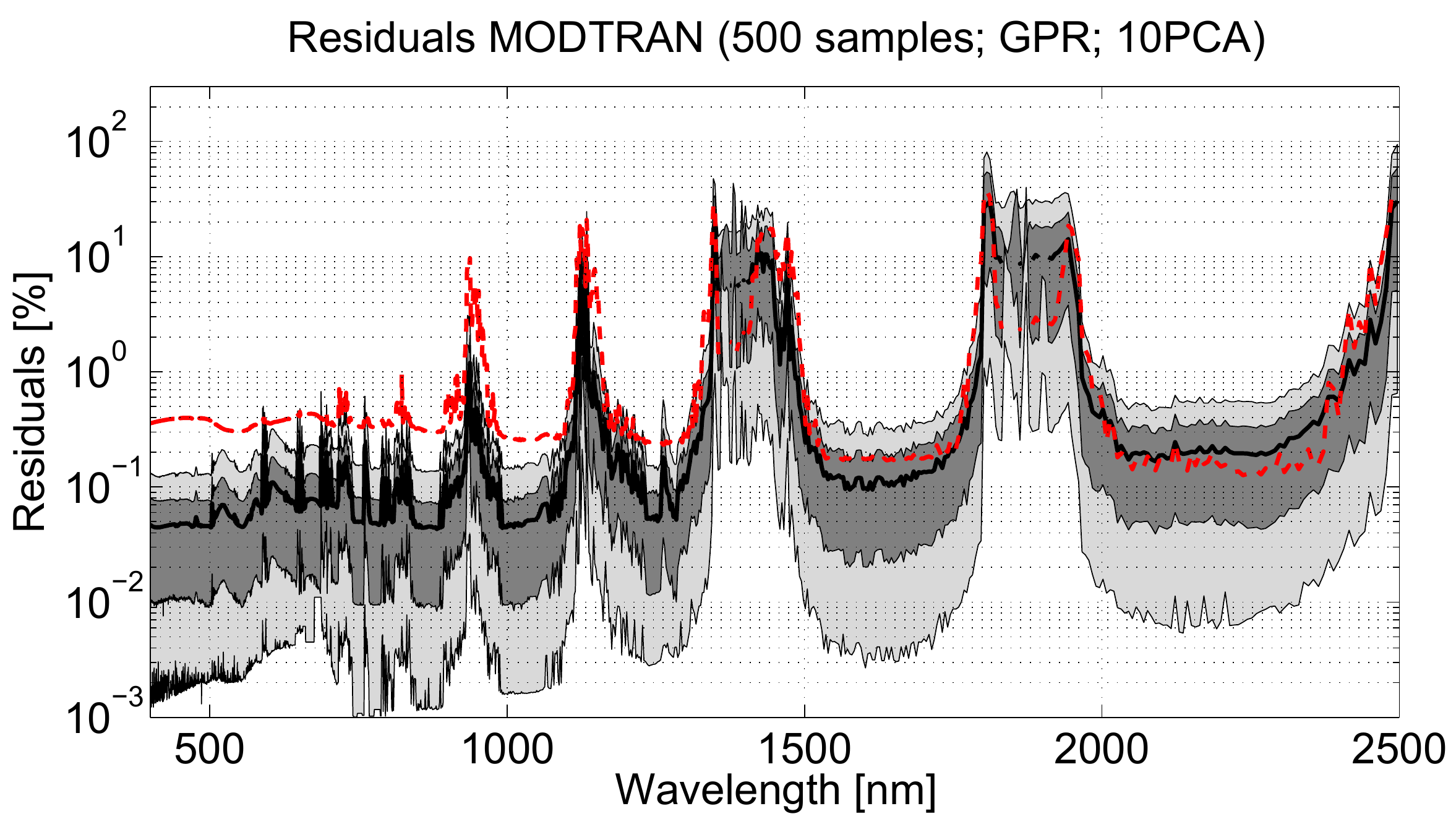}
	\IG[width=0.8\linewidth]{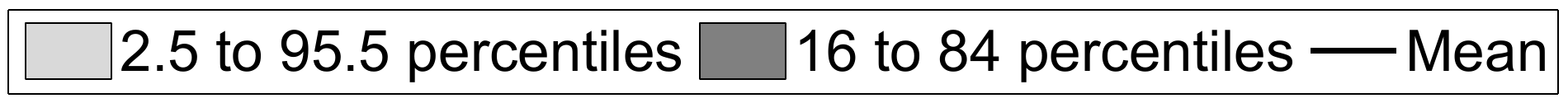}
	\caption{Histogram statistics of relative residuals (in absolute terms)(\%) for 2000-LUT linear interpolation (top) and 500-LUT GPR (bottom). The mean residual for linear interpolation is added on top of the GPR residuals (red dashed line).}
	\label{residualsMODTRAN}
\end{figure}

We now inspect the emulation/interpolation results in more detail through an analysis of the histograms of the relative residuals (in absolute terms) (see Figure \ref{residualsMODTRAN}). Linear interpolation method is shown as obtained with a 2000-LUT, whereas GPR is shown as trained with only a 500-LUT and 10 PCA components. Interestingly, although the emulator method is not presented in its optimized configuration, already a substantial gain in accuracy as compared to the optimized linear interpolation method is achieved. 

As previously observed in NRMSE values (see Table \ref{MODTRAN}), the reconstruction errors with GPR emulators obtains the best performance. The residual errors are in this case a factor 2-10 lower than using linear interpolation for both the lowest and highest errors in the histogram and in most part of the spectrum ($<$1800~nm).

\section{Conclusions}
\label{sec:conclusions}

E2E simulation of Earth Observation satellite missions rely on the execution of computationally expensive RTMs. Because these RTMs take long processing time, LUT interpolation techniques are typically used to sample their input variable space. However, the question arose whether the use of statistical methods, also known as emulators, can be used as a faster and more accurate altenative to classical interpolation techniques for E2E simulation applications. In this paper, we evaluated the accuracy and processing time of piece-wise linear interpolation and GPR techniques for to reconstruct TOA radiance data as from MODTRAN-based LUTs. The results showed that 
(1) GPR emulation produced output spectra with an error up to ten times lower than classical linear interpolation and 
(2) this with a faster processing speed (few seconds). 
It is thus concluded that GPR emulation offer a better alternative in computational cost and accuracy than traditional linear LUT interpolation, and therefore it opens new opportunities to improve the performance of E2E mission simulators.

Current work is carried out to include GPR emulators as an alternative to the current LUT interpolation methods implemented in the FLEX end-to-end mission simulator \cite{Vicent2016}. This will likely reduce the computation time for the generation of synthetic scenes, which will extend the current FLEX simulator capabilities to perform sensitivity analysis for various leaf/canopy and atmospheric conditions.


\begin{thebibliography}{1}

\bibitem{Verstraete2015419}
Verstraete. M.M, D.J. Diner, and J.-L. B\'ezy,
\newblock ``{Planning for a spaceborne Earth Observation mission: From user
  expectations to measurement requirements},''
\newblock {\em Environmental Science \& Policy}, vol. 54, pp. 419 -- 427, 2015.

\bibitem{kerekes1}
J.P. Kerekes and D.A. Landgrebe,
\newblock ``Simulation of optical remote sensing systems,''
\newblock {\em IEEE Transactions on Geoscience and Remote Sensing}, vol. 27,
  no. 6, pp. 762--771, 1989.

\bibitem{Tenjo2017}
C.~Tenjo, J.P. Rivera, N.~Sabater, J.~Vicent, L.~Alonso, J.~Verrelst, and
  J.~Moreno,
\newblock ``{Design of a generic 3D Scene Generator for Passive Optical
  Missions and its Implementation for the ESA's FLEX/Sentinel-3 Tandem
  Mission},''
\newblock {\em IEEE Transactions on Geoscience and Remote Sensing}, vol. 55,
  no. 13, pp. pp--pp, 2017.

\bibitem{guanter2009}
Luis Guanter, Rudolf Richter, and Hermann Kaufmann,
\newblock ``{On the application of the MODTRAN4 atmospheric radiative transfer
  code to optical remote sensing},''
\newblock {\em International Journal of Remote Sensing}, vol. 30, no. 6, pp.
  1407--1424, 2009.

\bibitem{OHagan2006}
A.~O'Hagan,
\newblock ``Bayesian analysis of computer code outputs: A tutorial,''
\newblock {\em Reliability Engineering and System Safety}, vol. 91, no. 10-11,
  pp. 1290--1300, 2006.

\bibitem{Berk2006}
A.~Berk, G.P. Anderson, P.K. Acharya, L.S. Bernstein, L.~Muratov, J.~Lee,
  M.~Fox, S.M. Adler-Golden, J.H. Chetwynd, M.L. Hoke, R.B. Lockwood, J.A.
  Gardner, T.W. Cooley, C.C. Borel, P.E. Lewis, and E.P. Shettle,
\newblock ``{MODTRAN}\textsuperscript{TM}5: 2006 update,''
\newblock 2006, vol. 6233 II.

\bibitem{Barber1996469}
C.B. Barber, D.P. Dobkin, and H.~Huhdanpaa,
\newblock ``The quickhull algorithm for convex hulls,''
\newblock {\em ACM Transactions on Mathematical Software}, vol. 22, no. 4, pp.
  469--483, 1996.

\bibitem{Verrelst2017}
J.~Verrelst, J.P. Rivera~Caicedo, J.~Mu\~{n}oz Mar\'i, G.~Camps-Valls, and
  J.~Moreno,
\newblock ``{SCOPE-based emulators for fast generation of synthetic canopy
  reflectance and sun-induced fluorescence Spectra},''
\newblock {\em Remote Sensing}, vol. 9, no. 9, 2017.

\bibitem{Vicent2016}
J.~Vicent, N.~Sabater, C.~Tenjo, J.R. Acarreta, M.~Manzano, J.P. Rivera,
  P.~Jurado, R.~Franco, L.~Alonso, J.~Verrelst, and J.~Moreno,
\newblock ``{FLEX End-to-End Mission Performance Simulator},''
\newblock {\em IEEE Transactions on Geoscience and Remote Sensing}, vol. 54,
  no. 7, pp. 4215--4223, 2016.

\end{thebibliography}
\end{document}